\documentclass[twocolumn]{autart}

\usepackage{graphicx,pstricks,scalerel}
\usepackage{cite}
\usepackage{amsmath,amssymb,amsfonts}
\usepackage{bm}

\makeatletter
\def\@xnamedef#1{\expandafter\protected@xdef\csname #1\endcsname}
\def\no@harm{} 
\def\ead@au#1{\protected@edef\@ead@au{#1}}
\patchcmd\runningauthor@fmt{\global\edef}{\protected@xdef}{}{}
\patchcmd\runningauthor@fmt{\global\edef}{\protected@xdef}{}{}
\patchcmd\author@fmt{\edef}{\protected@edef}{}{}
\patchcmd\add@xtok{\xdef}{\protected@xdef}{}{}
\makeatother

\newtheorem{theorem}{Theorem}
\newcommand{\mean}[1]{\mathcal{E}\left\{ #1 \right\}}
\newcommand{\abs}[1]{\left| #1 \right|}
\newcommand{\norm}[1]{\left|\left| #1 \right|\right|}
\newtheorem{lemma}[theorem]{Lemma}

\newtheorem{corollary}[theorem]{Corollary}
\newtheorem{assumption}[theorem]{Assumption}
\newcommand{\fin}{\hspace*{\fill}{$\square$}}

\usepackage{graphicx}          

\begin{document}
\begin{frontmatter}

 \title{On stochastic string stability with applications to platooning over additive noise channels  \thanksref{footnoteinfo}}

\thanks[footnoteinfo]{This work was supported by the Chilean National Agency for Research and Development (ANID) through the FONDECYT grant 1241813, 11221365, the FONDECYT Postdoctoral Grant 3230056, and the scholarship program ``Doctorado Nacional/2020-21202404''. The material in this paper was not presented at any conference. Corresponding author M.~A.~Gordon.}

\author[First]{Francisco J. Vargas} \ead{francisco.vargasp@usm.cl},
\author[First]{Marco A. Gordon}\ead{marco.gordon@sansano.usm.cl},
\author[Second]{Andr\'es A. Peters}\ead{andres.peters@uai.cl},  
\author[Third]{Alejandro I. Maass}\ead{alejandro.maass@uoh.cl}

\address[First]{Electronic Engineering Department, Universidad Técnica Federico Santa María, 2390123, Valparaíso, Chile}
\address[Second]{Faculty of Engineering and Sciences, Universidad Adolfo Ibáñez, Peñalol\'en, 7941169, Santiago, Chile}
\address[Third]{Institute
of Engineering Sciences, Universidad de O’Higgins, Rancagua, 2841959, Chile}

\begin{abstract}
This paper addresses the string stabilization of vehicular platooning when stochastic phenomena are inherent in inter-vehicle communication. To achieve this, we first provide two definitions to analytically assess the string stability in stochastic scenarios, considering the mean and variance of tracking errors as the platoon size grows. Subsequently, we analytically derive necessary and sufficient conditions to achieve this notion of string stability in predecessor-following linear platoons that communicate through additive white noise channels. We conclude that the condition ensuring string stability with ideal communication is essentially the same that achieves stochastic string stability when additive noise channels are in place and guarantees that the tracking error means and variances converge.
\end{abstract}	
\begin{keyword}                          
Stochastic string stability, Mean square string stability, Vehicular platooning, Additive white noise channels.
\end{keyword}                       
\end{frontmatter}

\section{Introduction}
Vehicle platooning control 
has gained great interest from the research community given the advances in autonomous navigation in recent years. In these applications, autonomous vehicles aim to travel coordinately, mostly employing information shared through wireless communication, and spacing coordination methods \cite{wang2020survey}. 
The main control problems in platooning consist of achieving the desired inter-vehicle distances and also ensuring \textit{string stability}, which is a property that guarantees that the detrimental effect of disturbances is limited as they propagate along the string of vehicles.
Due to its relevance, string stability has been widely studied for several platooning configurations. However, the majority of technical results related to this property rely on the assumption that communication between vehicles is not affected by random issues (see e.g. \cite{feng2019string} and the references therein). In practice, however, it is expected that inter-vehicle communications suffer random degradation, such as random data loss, and random delays, among others, which would require incorporating a stochastic approach for platooning analysis. 

 While string stability for deterministic systems has several definitions \cite{feng2019string} and has been extensively studied (see, for instance, \cite{peters2014,liu2021internal,ploeg2014lp,besselink2017string}), there are currently few analytical results providing conditions that tackle string stability in a stochastic setting. 
 We believe this is mainly due to the notion of stochastic string stability not being thoroughly explored yet (see \cite{socha2004,rybarska2007string} for early technical treatments of this topic). Nevertheless, there is a growing interest in studying platooning under random communication issues. 
For instance, the effect of random delays in platooning has been studied in \cite{zhao2021vehicle,elahi2022distributed,qin2017stability,liu2019robust,ma2020distributed,liu2021internal}, where different frameworks and solutions are proposed. However, for the aforementioned works, only  \cite{qin2017stability} presents an analytical definition for string stability considering the stochastic nature of the problem, which is named  $n\sigma$ string stability.
This notion aims to ensure that the trajectories within a neighborhood of the mean are string stable.
Platooning subject to random data loss has also received significant attention in the literature, where the detrimental impact of data loss has been reported, as well as the need to incorporate compensation strategies to deal with losses  
\cite{vargas2018,govape21,vivape23,li2019string,zhao2021stability,acciani2022stochastic,vegamoor2022string,gordon2023mean,rezaee2024cooperative}.  Despite the important results presented in these works, a limited number of them adopt an analytical approach to study stochastic string stability.
In \cite{li2019string}, an $\mathcal{L}_2$-norm-based string stability definition is presented and used to design an event-triggered scheme to deal with unreliable communication, including data loss.
The authors in \cite{zhao2021stability}  study both mean square stability and string stability for platooning with random packet drops. An analytical definition of string stability is given in \cite{zhao2021stability} in terms of the second moment of the signals of interest. 
The control design proposed in \cite{acciani2022stochastic} minimizes the error variance and achieves string stability defined in terms of the expected trajectory of the platoon. {More recently, \cite{rezaee2024cooperative} proposes a specific adaptive control strategy which ensures \emph{almost surely $\mathcal{L}_{\infty}$ string stability} under packet dropouts.}
Exploring additive white noise (AWN) channels in platooning, except for the initial exploration in \cite{gordon2020platoon}, is a novel area. This is surprising, given the importance of these channels in wireless communication analysis \cite{goldsmith2005}, and their substantial impact on control problems, extensively studied in works like \cite{vasichen2013,govachen2019,feng2023robust}.
Notably, the string stability notions in \cite{li2019string} and \cite{zhao2021stability} are not applicable to the AWN channels considered here; see Section \ref{SEC:m3s} for details.

 In this article, we aim to close part of the gap regarding the analytical treatment of string stability in stochastic settings, by providing technical definitions valid for a wide range of platooning configurations. We also apply these notions to the analysis of platooning over additive noise channels. This completes the string stability analysis exposed in our preliminary work \cite{gordon2020platoon}, where only a numerical study was carried out.   
The main contributions of this work can be summarized as follows:

    (1)\ We introduce the concept of  \textit{$\mathcal{L}_p$-mean $\mathcal{L}_q$-variance  string stability}, and also \textit{mean square string stability}.  These definitions capture the essence of the deterministic notion of $\mathcal{L}_p$ string stability \cite{ploeg2014lp} but are intricately linked to the well-established concept of \textit{mean square stability}, which is common in stochastic stability analysis \cite{soderstrom2002discrete}. These definitions apply to a wide range of platooning problems involving stochastic phenomena.\\
    (2)\ We apply the proposed concept of stochastic string stability to a homogeneous platoon with a predecessor-following topology, where inter-vehicle communication occurs through additive white noise channels. We analytically derive necessary and sufficient conditions for string stability in this setting. Our results show that the obtained condition are almost the same condition for string stability in the case of perfect communication. 

The present paper is arranged as follows. Section \ref{SEC:notation} defines notation and preliminaries. Section \ref{SEC:platoon_Setup} describes the platooning setup with additive noise channels.
Section \ref{SEC:m3s} presents string stability definitions suitable for stochastic settings. In Section \ref{SEC:m3s_additive_noise} the string stability of the platoon under study is analyzed. Simulation results are in Section \ref{SEC:example}. Lastly, Section \ref{SEC:conclusion} provides conclusions, while mathematical proofs are in the appendix sections.

\section{Notation and preliminaries}
\label{SEC:notation}
Let $\mathbb{N}_0$ be the set of natural number including zero ($\mathbb{N}_0\triangleq \mathbb{N}\cup \{0\}$), $\mathbb{R}_{\geq 0}\triangleq [0,\infty)$, and $\mathbb{R}^{m\times m}$ be the set of real matrices of dimension $m\times m$. A function $\gamma: \mathbb{R}_{\geq 0} \to \mathbb{R}_{\geq 0}$ is of
class-$\mathcal{K}$, if it is continuous, it vanishes at zero and is strictly increasing. 
Given a matrix $M \in \mathbb{R}^{m\times m}$, we denote its spectral radius by $\rho (M)$
its $i$-th singular value  by $\sigma_i(M)$, and its largest singular value $\sigma_{\max}(M)$. To denote the transpose, Kronecker product, and vectorization of a matrix, we use $(\cdot)^{\top}$, $\otimes$, and $vec(\cdot)$, respectively. Also, we utilize $vec^{-1}$ to denote the inverse operation of $vec$, where $M = vec^{-1} (vec (M))$. 
Consider a positive semi-definitive symmetric matrix $M$ and a column vector $x$, then the following properties hold \cite{bernstein2009matrix}: (a) $x^{\top}M x \leq \rho(M) \;x^{\top}x$, and (b)~$\sigma_{\max}(M)=\rho (M)$.
For any $x \in \mathbb{R}^{n}$, its  $p$-norm is defined as
$\norm{x}_p\triangleq \left(\sum_{i=1}^n \abs{x\{i \}}^p \right)^{1/p}$, for 
    $p \in \mathbb{N}$, where $x\{i \}$ denotes the $i$-th element in $x$. For a vector signal $x\in\mathbb{R}^n$, 
    $x(k)$  denotes the vector at time instant $k \in \mathbb{N}_{0}$.
    Its $\mathcal{L}_p$ norm is given by $\norm{x}_{\mathcal{L}_p}\triangleq \big(\sum_{k=0}^\infty \norm{x(k)}_p^p \big)^{1/p}$, for $p\in\mathbb{N}$, and $\norm{x}_{\mathcal{L}_\infty}\triangleq \displaystyle\sup_{k}  \norm{x(k)}_{\infty}$. For a constant, positive semi-definite, and symmetric  matrix  $X \in \mathbb{R}^{n \times n}$, we denote the Schatten norm (or $\sigma p$-norm) \cite{bernstein2009matrix} as 
$\norm{X}_{p}\triangleq \left(\sum_{i=1}^n \sigma_i(X)^p \right)^{1/p}$, for $p\in\mathbb{N}$, and $\norm{X}_\infty \triangleq \sigma_{\max}(X)$.
Lastly, for a time-variant matrix $X(k)$, $k\in\mathbb{N}_0$, we write $\norm{X}_{\mathcal{L}_{\infty}}\triangleq \displaystyle\sup_{k}  \norm{X(k)}_{\infty}$.\\
Let $\mean{\cdot}$ be the expectation operator. For a vector-valued discrete-time stochastic process $x$, we define the mean as $\mu_x(k) \triangleq \mean{x(k)}$, and the variance is determined by $P_x(k) \triangleq \mean{ (x(k)-\mu_x(k)) (x(k)-\mu_x(k))^\top}$, and the limiting values, if exist, are denoted by  $\mu^{\infty}_x\triangleq \displaystyle\lim_{k \rightarrow \infty} \mu_x(k)$, and $P^{\infty}_x\triangleq \displaystyle\lim_{k \rightarrow \infty} P_x(k)$.
A stochastic process $x$ is said to be a \emph{second order} process if and only if its mean and second moment matrix exist and are finite for all $k\in\mathbb{N}_{ 0}$.  
Consider a discrete-time LTI system with impulse response denoted by $w(k)$ and transfer function $W(z)$.  The output of such a system due to a stochastic process input $u$ can be obtained as $y(k)=w(k)*u(k)$, where $*$ denotes convolution. To ease notation, we will simply use $y=W u$. We also define $W^{\sim}(z) \triangleq W^{\top}(z^{-1})$. Finally, we use $W^{*}(e^{j\omega})$ to denote the complex conjugate transpose of $W(e^{j\omega})$. If $W(z)$ is a stable SISO transfer function, its $\mathcal{H}_2$ norm and $\mathcal{H}_\infty$ norm are given by $\| W \|_2 = \left( \frac{1}{2\pi} \int_{0}^{2\pi} | W(e^{j\omega})|^{2}d\omega \right)^{1/2} $, and $\| W\|_{\infty} = \displaystyle\max_{\omega \in [0,2\pi ]} |W(e^{j\omega}) |$, respectively.

\section{Platoon over additive noise channels}
\label{SEC:platoon_Setup}

\begin{figure}
		\begin{center}
			\includegraphics[width=0.9\columnwidth]{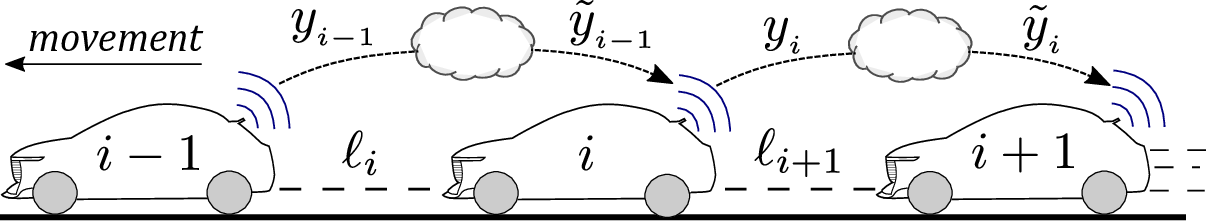}
			\caption{Platoon configuration}
			\label{Fig:platoon}
		\end{center}
	\end{figure}

We consider $N\in \mathbb{N}$ autonomous vehicles, also called followers, and a leading vehicle. All of them are moving in a one-dimensional track as shown in Fig.~\ref{Fig:platoon}. Each vehicle is meant to follow its corresponding predecessor while maintaining a desired inter-vehicle distance using wireless communication.  Every vehicle in the platoon is identified by its corresponding location in the string with the sub-index $i$, where $i=0,1,2,\dots,N$. The index $i=0$ corresponds to the leading vehicle, while the index $N$ corresponds to the last follower in the string. 
 We focus on platoons with homogeneous vehicles, i.e., we assume all the agents have the same dynamics and controller. 
 In this paper, we will adopt a discrete-time setting to study vehicle platooning, which is a natural consideration motivated by digital wireless communication and by the fact that most controllers nowadays are implemented in digital devices. 
Certainly, the inherent nature of the vehicle dynamics is in continuous time, nevertheless, a discrete-time model may be obtained through the application of standard sample-data model techniques \cite{aastrom2013computer}. Moreover, we consider a predecessor-following topology, in which each vehicle has access to its own position and velocity, but also transmits its position to its immediate follower. 
 
At any instant time $k \in \mathbb{N}_0$, the inter-vehicle distance between the $i$-th and $(i-1)$-th agents is defined as 
\begin{equation*}
    \ell_i(k) \triangleq y_{i-1}(k) - y_i(k),
\end{equation*}
where $y_i(k)$ denotes the position of the $i$-th vehicle. 	The control objective is to maintain the inter-vehicle distance $\ell_i(k)$ as close as possible to a desired reference $r_i(k)$.
However, we consider that the measured position by the $(i-1)$-th vehicle is sent through a wireless additive noise channel to the immediate follower in the $i$-th location. As a consequence, the $i$-th vehicle only has access to a noisy measurement of the predecessor's position. The received position is given by $\tilde{y}_{i-1}(k)=y_{i-1}(k)+d_i(k),$ where $d_i$ is a stochastic process modelling channel noise. Note that, the $i$-th vehicle cannot calculate $\ell_i(k)$. Instead, a corrupted version (denoted by $\tilde \ell_i(k)$) is used by the local control loop, which is defined as
$\tilde \ell_i(k) \triangleq \tilde{y}_{i-1}(k)-y_i(k)$. 

\begin{figure}[t]
    \begin{center}
    \includegraphics[width=0.8\columnwidth]{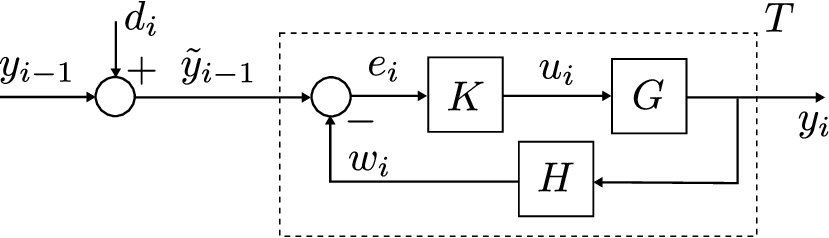}
    \caption{Feedback control loop for the $i$-th agent}
    \label{fig:closed_loop}                   
    \end{center}                             
\end{figure}

We adopt a constant time headway spacing policy, which yields a desired spacing between two vehicles that increases/decreases along with the speed at which a vehicle approaches its predecessor \cite{ploeg2014lp}. Therefore, the reference distance  $r_i(k)$ is not constant as it depends on the velocity of each agent. 
With this policy, we have that
\begin{equation}
    \label{Ec:policy} 
    r_i(k)=\delta_i+h \left[ y_i(k)-y_i(k-1) \right],
\end{equation}
where $\delta_i \geq 0$ represents the minimum desired distance between agents, $y_i(k)-y_i(k-1)$ represents the velocity of the $i$-th agent\footnote{The discrete-time representation of the velocity is given by $(y_i(k)-y_i(k-1))/T_s$, where $T_s$ denotes the sample time. However, w.l.o.g., the term $T_s$ can be absorbed by $h$.}, and $h>0$ is the time headway constant. For ease of exposition, we assume the length of the vehicles is equal to zero and the scenario where $\delta_i=0$. Since $\delta_i$ is a constant and the setup is linear, this assumption does not impact the subsequent string stability analysis.  Consequently, the position error $\zeta_i(k) \triangleq \ell_i(k)-r_i(k)$ can be calculated as 
\begin{equation} 
    \label{Ec.error1}
    \zeta_i(k)=y_{i-1}(k)-y_i(k)- h\left[ y_i(k)-y_i(k-1) \right],
\end{equation}
while the error in the local control loop is given by
\begin{equation} \label{eq:error_f_de_zeta}
    e_i(k)=\tilde \ell_i(k)-r_i(k)
    =\zeta_i(k)+d_i(k).
\end{equation}
The dynamics of each vehicle are modeled as an LTI system $G(z)$, which is locally controlled by an LTI controller $K(z)$. The feedback control system in each vehicle can be described as in Fig. \ref{fig:closed_loop}, where $T(z)$ denotes the closed-loop transfer function, and $H(z)$ incorporates the time headway policy in the feedback path of the closed-loop. Both transfer functions are given by
\begin{equation*}
    T(z)=\dfrac{K(z)G(z)}{1+K(z)G(z)H(z)},\; \; \ H(z)=(1+h)-\dfrac{h}{z}. 
\end{equation*}
The transfer function from $y_{i-1}$ to the error $e_i$, is the sensitivity $S(z)$, which satisfies $S(z)=1-H(z)T(z)$.
We adopt the following assumption for the vehicle model and the additive noises.

 	\begin{assumption}\label{assum:tf_noise}
  \hspace{0cm}
        \begin{enumerate}
            \item[a)] $T(z)$ is strictly proper (has more poles than zeros).
            \item[b)] The product $G(z)K(z)$ has, at least, double integral action (at least two poles at $z=1$)
            \item[c)] The noises $d_i$ are second-order, stationary, and mutually independent processes.
            \item[d)] The noises $d_i$ are identically distributed white noise processes, with mean $\mu_{d_i}=0$ and variance  $P_{d_i}=P_d$.  
            \item[e)] The noises $d_i(k)$ affect the platoon for $k \geq 0$, and are uncorrelated with the initial state of all vehicles.      
         \end{enumerate}
    \end{assumption} 
    Assumption \ref{assum:tf_noise} a) and b) are required to have a well-defined control loop able to track ramp signals. Note that $T(1)=1$ is a consequence of b). Assumptions c), d) and e) are common in control problems to model multiple additive noise channels\cite{vasichen2013}.
In a deterministic setting with ideal communication, it would be natural to also assume that $T(z)$ is internally stable, achieving zero error in steady state, and such that $\norm{T(z)}_\infty \leq 1$ to allow for string stability \cite{ploeg2014lp,vargas2018}; however, such an assumption is not adopted \emph{a priori} here.

We use a minimal state-space representation for the $i$-th ($i \geq 1$) closed-loop system $T$ given by
\begin{subequations}\label{eq:ith-closedloop}
    \begin{align}
	     x_i(k+1)&=A x_i(k) + B y_{i-1}(k)+Bd_i(k), \\
	     y_i(k)&=C x_i(k), \quad k \in \mathbb{N}_0
    \end{align}
\end{subequations}
where $x_i \in \mathbb{R}^{n}$ with initial conditions $x_i(0)$, $A\in \mathbb{R}^{n \times n}$, $B \in \mathbb{R}^{n \times 1}$, and $C \in \mathbb{R}^{1 \times n}$. The state vector $x_i$ typically contains position, velocity, acceleration, and potentially additional variables, which depend on the model $G$ and the controller $K$. Note that the state-space representation of $T$  is not confined to a specific type of controller, nor is it limited to a particular dynamic model. Any plant and controller that satisfy Assumption \ref{assum:tf_noise} are admissible in our setup. 

\begin{figure}[t]
    \begin{center}
    \includegraphics[width=0.99\columnwidth]{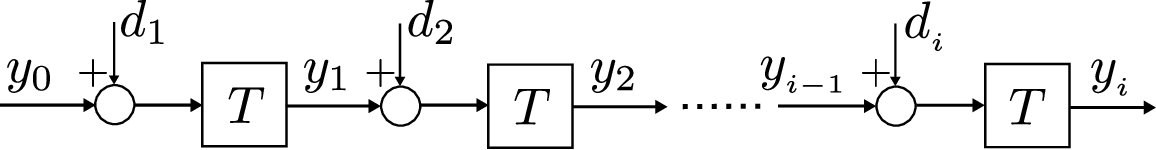}
    \caption{Concatenation of the platoon with channel noise on each vehicle.} 
    \label{fig:concatenacion_pert_all}
    \end{center}                             
\end{figure}

The leader trajectory $y_0$ is a reference for the remaining vehicles, which is depicted in Fig. \ref{fig:concatenacion_pert_all}.
It is expected that the leader maintains predominantly a cruising speed, resulting in the leader's position exhibiting a characteristic ramp signal. To avoid such unbounded signals in our setup, we model 
the movement of the leader as the consequence of tracking a virtual leader \cite{ploeg2014lp,yu2021automated,elahi2022distributed}. This approach allows for the consideration of the leader's dynamics in the movement of the platoon, as well as considering bounded signals as inputs to the platoon.

\begin{assumption}\label{assum:leader}
  \hspace{0cm}
        \begin{enumerate}
        \item[a)] The leader ($i=0$)
follows a virtual reference $r_0$, and it is not affected by noise.
\item[b)]  The leader trajectory $y_0$ becomes a ramp signal (cruise velocity) when tracking $r_0$, with sporadic changes due to external disturbances or changes in the leader's speed.
\item[c)] The tracking error of the leader when tracking the virtual reference $r_0$, denoted by $\zeta_0$, is deterministic and converges to zero when $k$ goes to infinity. \fin 
\end{enumerate}
    \end{assumption} 



Since $y_0$ is expected to converge to a ramp signal, we can write an alternative platoon representation based on $\zeta_0$  as the main deterministic input. Given that the leader is not affected by noise, we have that (see \eqref{Ec.error1}) $\zeta_0(k)=e_0(k)=r_0(k)-y_0(k)- h\left[ y_0(k)-y_0(k-1) \right]$. 
Note that, given Assumption \ref{assum:leader}.c), the power spectral density of $\zeta_0$ is zero.

In platooning problems, vehicle positions $y_i(k)$ are expected to converge to a ramp signal when cruise velocity is reached. Therefore, some convergence properties may not be applicable to a state space representation based on the vehicle states $x_i(k)$. Since we are interested in studying the convergence of the error signal, we define an alternative state $\xi_i(k)$ that is expected to converge to an equilibrium point and describe the tracking errors in terms of such state (instead of $x_i$). So we define
\begin{equation}\label{eq:statesxi}
    \xi_i(k)\triangleq x_{i-1}(k)-x_i(k)- h\left[ x_i(k)-x_i(k-1) \right], 
    \end{equation} 
which is composed of the differences $x_{i-1}(k) -  x_i(k)$ and $x_{i}(k) -  x_i(k-1)$. Such differences are expected to converge with a suitable controller design and ideal communication, even if some components in the states $x_i$ and $x_{i-1}$ become ramps. Consequently,  a state representation based on $\xi_i$ eases the study of convergence in the mean square sense.

To analyze the overall platoon, we can concatenate the signals of interest in each vehicle and define the state $\bm{\xi}(k) \triangleq \begin{bmatrix} \xi_1(k)^{\top} & \xi_2(k)^{\top} & \dots & \xi_N(k)^{\top} \end{bmatrix} ^{\top}$, and the error vector $\bm{\zeta}(k) \triangleq  \begin{bmatrix} \zeta_1(k) & \zeta_2(k) & \dots & \zeta_N(k) \end{bmatrix} ^{\top}$. Utilizing the definitions in \eqref{eq:statesxi}, and \eqref{Ec.error1}, it is not difficult to write the system in \eqref{eq:ith-closedloop} as a function of the alternative state as 
\begin{subequations} \label{eq:ss_sys_all}
    \begin{align}
        \label{eq:xi_all}
        \bm{\xi}(k+1) &=\mathbf{A} \bm{\xi}(k) + \mathbf{B}_{o}\, \zeta_0(k)+\mathbf{B}_{a}\, \mathbf{d}(k) +\mathbf{B}_{b}\, \mathbf{d}(k-1) \\
        \label{eq:z_all}
        \bm{\zeta}(k) &= \mathbf{C} \bm{\xi}(k),
    \end{align}
\end{subequations}
where $\mathbf{d}(k) \triangleq  \begin{bmatrix} d_1(k) & d_2(k) & \dots & d_N(k) \end{bmatrix} ^{\top}$, $\mathbf{B}_{b}\triangleq h \mathbf{B} $, $\mathbf{B} \triangleq  \text{diag}\begin{bmatrix}
        B & \cdots &  B
        \end{bmatrix}$, $ \mathbf{B}_{o} \triangleq  \begin{bmatrix}
        B^\top & 0 & 0 & \cdots &  0
        \end{bmatrix}^{\top}$,
        $ \mathbf{C} \triangleq  \text{diag}\begin{bmatrix}
        C & \cdots &  C
        \end{bmatrix},$ and considering $\lambda\triangleq -(1+h)$ we obtain
\begin{align*}
    \mathbf{A}&\triangleq 
        \begin{bmatrix}
            A & \ & \ & \ \\
            BC & A & \ & \ \\
            \ & \ddots & \ddots & \ \\
            \ & \ & BC & A
        \end{bmatrix}, \text{ and}\;
    \mathbf{B}_{a}&\triangleq \begin{bmatrix}
            \lambda B &   &  \  &  \  \\
            B &  \lambda B &    \  &  \  \\
            & \ddots  &  \ddots & \ \\
            &  & B & \lambda B
    \end{bmatrix}.
\end{align*}
It is assumed that $\bm{\xi}(0)\in \mathbb{R}^{nN}$ is a second-order random variable with a positive definite covariance matrix.\\
On the other hand, it is also useful to have an iterative description of the tracking errors in the platoon in \eqref{eq:ss_sys_all}. Given the structure of the state space matrices in \eqref{eq:ss_sys_all}, the relation between $e_i$ and $\zeta_i$ in \eqref{eq:error_f_de_zeta} and $S(z)=1-H(z)T(z)$, it is not difficult to show that
\begin{align}
    \label{eq:mu_z_recursive_all}
    \zeta_i & =
    \begin{cases}
        T \zeta_0 - HT d_1, & \text{for } i=1, \\
        T \zeta_{i-1} + T d_{i-1} - HT d_i, & \text{for } i>1.
    \end{cases}
\end{align}
where we used a simplified notation as per Section \ref{SEC:notation}.

\section{String stability for stochastic systems} \label{SEC:m3s}

String stability studies the boundedness of disturbances as they propagate through a string of interconnected vehicles. While an array of definitions exists in \cite{feng2019string}, they are designed for deterministic settings and are not directly applicable to platoons with stochastic processes, as seen in our communication channel model. Addressing this gap, our work focuses on stochastic settings, an area less explored in current literature.

Three relevant signals to study string stability are the tracking errors $\bm{\zeta}$, the state of the platoon $\bm{\xi}$, and a vector of exogenous inputs denoted by $\bm{\nu}$, which are stochastic process in our setup. In general, $\bm{\nu}$ may contain the desired trajectory of the leader, external disturbances, and noises. Additionally, the initial condition $\bm{\xi}(0)$ is also relevant for string stability analysis. In our setting, the vector of exogenous inputs can be set as $$\bm{\nu}(k) \triangleq \begin{bmatrix} \nu_1(k)^{\top} & \nu_2(k)^{\top} & \dots & \nu_N(k)^{\top} \end{bmatrix} ^{\top}=\eta_1 \zeta_0(k) +\mathbf{d}(k).$$
We denote as $\mathcal{G}$ a set of technical specifications required for the initial condition and exogenous input.  We set $\mathcal{G}$ as the set of initial condition $\bm{\xi}(0)$ and external inputs $\bm{\nu}$ such that $\norm{\mu_{\bm{\xi}}(0)}_p$, $\norm{P_{\bm{\xi}}(0)}_q$, $\norm{\mu_{\bm{\nu}}(k)}_{\mathcal{L}_p}$ and
$\norm{P_{\bm{\nu}}(k)}_{\mathcal{L}_q}$ are finite,  for some $p$ and $q$ in $\{1,2, \dots,\infty \}$. 


\begin{defn}\label{def:Lp_Lq_ss}
A platoon of vehicles is said to be $\mathcal{L}_p$-mean $\mathcal{L}_q$-variance  string stable
with  $p,q \in \mathbb{N}\cup\{+\infty\}$, if there exist class-$\mathcal{K}$ functions $\alpha_1$, $\alpha_2$,  $\beta_1$ and $\beta_2$, such that, 
for any $\{\bm{\xi}(0),\bm{\nu}\} \in \mathcal{G}$, the following holds
\begin{align*}
\norm{\mu_{\zeta_i}(k)}_{\mathcal{L}_p} &\leq \alpha_1(\norm{\mu_{\bm{\nu}}(k)}_{\mathcal{L}_p})+\beta_1(\norm{\mu_{\bm{\xi}}(0)}_p), \\
\norm{P_{\zeta_i}(k)}_{\mathcal{L}_q} &\leq \alpha_2 (\norm{P_{\bm{\nu}}(k)}_{\mathcal{L}_q})+\beta_2(\norm{P_{\bm{\xi}}(0)}_q), 
\end{align*}
 $\forall k\in\mathbb{N}_0$, $\forall i \in \left\lbrace 1, \dots, N\right\rbrace$ and $\forall N \in \mathbb{N}$.
\fin 
\end{defn}

Definition \ref{def:Lp_Lq_ss} ensures bounded second-order statistics for $\zeta_i$,  $\forall i \in \left\lbrace 1, \dots, N\right\rbrace$ and $\forall N \in \mathbb{N}$. These bounds are independent of $N$. This is crucial in string stability definitions, as it signifies establishing a property linked to the scalability of the interconnected system.
Definition \ref{def:Lp_Lq_ss} is inspired by the concept of deterministic \textit{$\mathcal{L}_p$ string stability} from \cite{ploeg2014lp} and adapted to stochastic settings, offering flexibility by explicitly considering the norm of the mean and variance separately. This stems from the fact that, depending on the underlying problem, certain signal norms may exist while others may not, as mentioned earlier.
Certainly, the string stability definitions in \cite{li2019string,zhao2021stability} rely on the $\mathcal{L}_2$ norm, making them inapplicable when second-order statistics fail to converge to zero.
Furthermore, compared to others, our definition enables the explicit detection of cases where the mean converges to stationary values but variances do not, as seen in \cite{gordon2023mean}. This is more practical and allows to  accommodate fluctuations within certain statistical bounds. In fact, Definition \ref{def:Lp_Lq_ss} is more general than \cite{acciani2022stochastic} (mean-only consideration), \cite{li2019string,zhao2021stability} ($\mathcal{L}_2$ norm focus), \cite{qin2017stability} (string stability only in the mean's neighborhood), and \cite{rezaee2024cooperative} (almost sure $\mathcal{L}_{\infty}$ string stability).

In Definition \ref{def:Lp_Lq_ss}, the choice of $p$ and $q$, naturally requires the existence of the corresponding norms. For instance, if $\lim_{k \rightarrow \infty} \mu_{\zeta_i}(k) \neq 0$, then
$\norm{\mu_{\zeta_i}(k)}_{\mathcal{L}_2}$ would not be bounded. 
We illustrate this fact in the numerical simulations from Section \ref{SEC:example}. 
Note that $\bm{\xi}(0)$ is a second-order random variable,  and hence the existence of the corresponding norms is guaranteed for $p$ and $q$ in $\{1,2,\infty \}$.
On the other hand, we have that 
$\mu_{\bm{\nu}}(k)=\eta_1 \zeta_0(k) $, and $P_{\bm{\nu}}(k) = P_{\bm{d}}(k) = P_d \; I^{N \times N} $ respectively. Thus, the corresponding $\mathcal{L}_2$ and $\mathcal{L}_\infty$ norms exist, and are such that $\norm{\mu_{\bm{\nu}}(k)}_{\mathcal{L}_2}=\norm{\zeta_0(k)}_{\mathcal{L}_2}$ and $\norm{P_{\bm{\nu}}(k)}_{\mathcal{L}_\infty}=P_{d}$. We highlight that, in this setup, we cannot study $\mathcal{L}_2$-variance string stability since $\norm{P_{\bm{\nu}}(k)}_{\mathcal{L}_2}=\infty$.

\begin{defn}\label{def:msss}
		An $\mathcal{L}_p$-mean $\mathcal{L}_q$-variance  string stable  platoon of vehicles is said to be mean square string stable if for any  $\{\bm{\xi}(0),\bm{\nu} \} \in \mathcal{G}$, there exist finite constants $M_1$ and $M_2$ such that
   $\displaystyle \lim_{i\to \infty}\lim_{k \to \infty}\mu_{\zeta_i}(k)  = M_1$, and $\displaystyle \lim_{i\to \infty}\lim_{k \to \infty}P_{\zeta_i}(k)  = M_2$.
  $\hfill\square$
\end{defn}

Definition \ref{def:msss} ensures convergence of the second order moment of the stochastic process involved, which is a standard requirement in stochastic control theory \cite{astrom2012,soderstrom2002discrete}. Indeed, a system is said to be mean square stable (MSS) if the mean and variance of the signal of interest converge. By merging this concept with the one outlined in Definition~\ref{def:Lp_Lq_ss}, we can articulate the proposed \emph{mean square string stability} notion.\\
We highlight that, similar to the deterministic case, the proposed concept of stochastic string stability does not guarantee collision avoidance but rather facilitates it.

\begin{rem}
It is important to highlight that these definitions are not confined to any specific type of vehicle model, control strategy, spacing policies, or communication topologies. Indeed, they are also applicable to platooning scenarios involving other random communication issues, such as random data loss, random delays, etc., provided that suitable definitions for the signals $\bm{\xi}$, $\bm{\nu}$, and the set $\mathcal{G}$ are specified within the respective frameworks.  Additionally, we would like to mention that analogous notions to Definitions \ref{def:Lp_Lq_ss} and \ref{def:msss} for continuous-time settings are not difficult to establish. 
\end{rem}

\section{String stability analysis for platoon under additive noise channels} \label{SEC:m3s_additive_noise}

We first study the properties of the second-order statistics of the tracking error as the time approaches infinity, then we consider an unlimited number of vehicles, and lastly, we present conditions for string stability. The mathematical proofs of the results in this section are located in the appendices section.
\subsection{Analysis for $k \rightarrow \infty$} \label{sec:time_all}
In the following Lemma, we present time-domain expressions for the mean and variance of the tracking errors and also provide a necessary and sufficient condition for mean square stability.

\begin{lemma} \label{lem:mean_square_all}
Consider the platoon described by \eqref{eq:ss_sys_all}. Then, the mean of the state and the tracking error are given by 
\begin{subequations} \label{eq:means_all}
\begin{align}
    \label{eq:mu_xi_all}
    \mu_{\bm{\xi}}(k+1) &= \mathbf{A} \mu_{\bm{\xi}}(k) +\mathbf{B}_o\ \zeta_0(k), \quad \mu_{\bm{\xi}}(0)=\mu_{\bm{\xi}_0}, \\
        \label{eq:mu_z_all}
    \mu_{\bm{\zeta}}(k) &=
        \mathbf{C} \mu_{\bm{\xi}}(k).
\end{align}	  
\end{subequations}
The corresponding covariance matrices are given by
\begin{subequations} \label{eq:variance_all}
\begin{align} 
    \label{eq:P_xi_all}
    P_{\bm{\xi}}(k+1) &=
    \begin{cases}
        \mathbf{A} P_{\bm{\xi}}(k)\mathbf{A}^\top + \mathbf{B}_a P_{\bm{d}} {\mathbf{B}_a}^\top,  & \text{for } k=0, \\
        \mathbf{A} P_{\bm{\xi}}(k)\mathbf{A}^\top + \Upsilon, & \text{for } k>0,
    \end{cases} \\
    \label{eq:P_z_all}
    P_{\bm{\zeta}}(k) &=
        \mathbf{C} P_{\bm{\xi}}(k)\mathbf{C}^\top,
\end{align}
\end{subequations}
where $P_{\bm{\xi}}(0) = P_{\bm{\xi}_0}$ and $\Upsilon = \mathbf{B}_{a} P_{\mathbf{d}} {\mathbf{B}_{a}}^\top + \mathbf{B}_{b} P_{\mathbf{d}} {\mathbf{B}_{b}}^\top + \mathbf{A}\mathbf{B}_{a} P_{\mathbf{d}} {\mathbf{B}_{b}}^\top + \mathbf{B}_{b} P_{\mathbf{d}} {\mathbf{B}_{a}}^\top \mathbf{A}^{\top}$. Additionally, $\mu_{\bm{\zeta}}(k)$ and $P_{\bm{\zeta}}(k)$ converge in time to stationary values, for all $N \in \mathbb{N}$, if and only if
    \begin{equation}
        \label{eq:MSS_all}
        \rho (A)<1.
    \end{equation}
   Said stationary values are given by
   $\mu^{\infty}_{\bm{\zeta}} = 0$ and $P^{\infty}_{\bm{\zeta}} = vec^{-1} \left( (\mathbf{C} \otimes \mathbf{C})  (I- \mathbf{A} \otimes \mathbf{A} )^{-1} vec(\Upsilon)  \right)$. 
\end{lemma}
 It is not difficult to show that $P^{\infty}_{\bm{\xi}}$ can be alternatively obtained from the Lyapunov equation $P^{\infty}_{\bm{\xi}}=\mathbf{A} P^{\infty}_{\bm{\xi}}\mathbf{A}^\top + \Upsilon$. In  Lemma \ref{lem:mean_square_all},  the time convergence condition \eqref{eq:MSS_all}  does not change with the platoon length, and the stationary value does not depend on the initial conditions.
Clearly,  \eqref{eq:MSS_all} ensures internal stability of $T(z)$ ---a basic requirement in the deterministic setting--- but also mean-square stability (MSS) due to  the second-order statistic convergence, which is a fundamental requirement when dealing with stochastic settings \cite{astrom2012}.
Furthermore, \eqref{eq:MSS_all} also ensures the existence of stationary statistics of the errors, including stationary power spectral densities. This fact is used in the following section to study the influence of platoon length.

\subsection{Analysis for $N \rightarrow \infty$} \label{sec:scalability_all}

 As the next step before providing string stability conditions, we show that the platoon stationary errors remain bounded when the number of vehicles grows unbounded. 
Assuming $\rho(A) < 1$, the time convergence of the system \eqref{eq:ss_sys_all} is guaranteed, which also ensures the existence of a well-defined stationary mean and power spectral density. The latter is given as a recursive expression in the frequency domain in the next theorem.

\begin{theorem} \label{lem:zdom_scalability_all}
   Consider the system described in \eqref{eq:ss_sys_all} and assume $\rho(A) < 1$. The tracking error of the $i$-th follower, with $i \in \lbrace 2,3,...,N\rbrace$, $N \in \mathbb{N}$, is such that its mean satisfies the recurrence relation
    \begin{align}
    \label{eq:zdom_muz_all}
    \mu_{\zeta_{i}}(e^{j\omega}) =& T(e^{j\omega}) \mu_{\zeta_{i-1}} (e^{j\omega}),
    \end{align}
    where $\mu_{\zeta_{1}}(e^{j\omega}) = T(e^{j\omega}) \zeta_{0} (e^{j\omega})$. 
    Additionally, each element $\mu_{\zeta_{i}}(e^{j\omega})$ is bounded $\forall \omega$, and remains bounded when $i \rightarrow \infty$, if and only if  $\norm{T(z)}_\infty \leq 1$.
    %
    Furthermore, its power spectral density satisfies 
    \begin{align}
     \label{eq:zdom_psdz_all}
    \phi_{\zeta_i}(e^{j\omega}) =&
        |T(e^{j\omega})|^2 \phi_{\zeta_{i-1}} (e^{j\omega}) + |T(e^{j\omega})|^2 \ |S(e^{j\omega})|^2 P_d \notag \\
        &+ \left( 1-|T(e^{j\omega} \right)|^2)\ |1-S(e^{j\omega})|^2  P_d,
    \end{align}
    where $\phi_{\zeta_1}(e^{j\omega})=|H(e^{j\omega})T(e^{j\omega})|^2 P_d$. 
    Moreover, each element $\phi_{\zeta_i}(e^{j\omega})$  is bounded $\forall \omega$, and remains bounded when $i \rightarrow \infty$, if and only if
        \begin{equation} \label{eq:scalability_all}
    |T(e^{j\omega})| <1, \quad \forall \omega >0 
\end{equation}
\end{theorem} 

Theorem \ref{lem:zdom_scalability_all} not only characterizes the mean and stationary spectrum of each vehicle, but it also provides  necessary and sufficient conditions to ensure they remain bounded as the number of vehicles goes to infinity. These conditions remain invariant regardless of the platoon length. 
In Theorem \ref{lem:zdom_scalability_all}, we note that $\norm{T(z)}_\infty \leq 1$ is a necessary and sufficient condition for the sequence $\{\mu_{\zeta_{i}}(e^{j\omega}) \}$  to be bounded, but not for
$\{\phi_{\zeta_i}(e^{j\omega}) \}$. In fact, the problem setup requires $|T(e^{j0})|=1$, but there may exist non-zero $\omega_{Q}$ such that $|T(e^{j\omega_{Q}})|=1$, and thus the recursion $\phi_{\zeta_N}(e^{j\omega_{Q}})$ is not bounded since $S(e^{j\omega})=0$ only for $\omega=0$. This explains the subtle difference between condition for $\mu_{\zeta_{i}}(e^{j\omega})$ and the one for $\phi_{\zeta_i}(e^{j\omega})$ in \eqref{eq:scalability_all}.
We also highlight that if $\rho(A)\geq 1$, it is not possible to characterize the power spectral density of the errors. In the context of controller design, well-established techniques from robust control for linear systems \cite{zhou1996robust} can be utilized to satisfy conditions in Theorem \ref{lem:zdom_scalability_all}. 





\begin{corollary}\label{cor:stationary_all}
Consider the platoon \eqref{eq:ss_sys_all} that satisfies conditions \eqref{eq:MSS_all} and \eqref{eq:scalability_all}. Then, for  $i \in \lbrace 1,2,...,N\rbrace$, $\forall N \in \mathbb{N}$, the stationary variances $P^{\infty}_{\zeta_i}$ satisfy
\begin{enumerate}
\item $P^{\infty}_{\zeta_{i-1}} \leq  P^{\infty}_{\zeta_i} \qquad \forall i>1$.
    \item $\displaystyle\max_{i} P^{\infty}_{\zeta_i} =  P^{\infty}_{\zeta_N}  =\|F_{\zeta_N}\|_{2}^2 \; P_d$.
\item $\displaystyle\lim_{N \rightarrow \infty} P^{\infty}_{\zeta_N}=\left( \norm{ \dfrac{S(z)}{M(z)} }_2^2 -1 \right) P_d .$
\end{enumerate}
 where $F_{\zeta_N} = \left[ ST^{N-1}\ \ ST^{N-2}\ \ \cdots \  ST \, \, -HT \right]$ and $M(z)$ is a stable and minimum phase spectral factor such that $1-T(z)T(z)^{\sim}=M(z)M(z)^{\sim}$.
\end{corollary}

Corollary \ref{cor:stationary_all} states that, if the aforementioned conditions are met, then the stationary variances increase with the position index $i$, and are upper bounded by a term that does not depend on $N$. 
This upper bound is stationary and not necessarily valid for any time $k$, since some  initial conditions may generate higher transient variances.

\subsection{String stability conditions} \label{sec:string_stability_all}
We are now in a position to provide string stability conditions for the platoon described by system \eqref{eq:ss_sys_all}. According to the findings in Theorem \ref{lem:zdom_scalability_all} and Corollary \ref{cor:stationary_all}, it is established that the mean of each vehicle converges to zero over time, whereas the variance of each vehicle converges to a non-zero stationary value. Based on these convergences, we provide conditions that ensure $\mathcal{L}_2$-mean $\mathcal{L}_\infty$-variance string stability as per Definition \ref{def:Lp_Lq_ss}; conditions that further ensure mean square string stability as per Definition \ref{def:msss} are also provided. It is important to highlight that, for our studied setting, we focus on ensuring $\mathcal{L}_2$-mean $\mathcal{L}_\infty$-variance string stability. However, Definition \ref{def:Lp_Lq_ss} is valid for any $p,q\in\mathbb{N}\cup\{+\infty\}$, provided the corresponding norms exist. Additional scalability properties could be shown for other pairings of $p,q$, but this is beyond the scope of this paper.


\begin{theorem} \label{teo:Lpmean_Lqvariance_all}
Consider the platoon described by \eqref{eq:ss_sys_all}. The conditions in  \eqref{eq:MSS_all} and \eqref{eq:scalability_all} are necessary and sufficient to achieve  $\mathcal{L}_2$-mean $\mathcal{L}_\infty$-variance string stability, but also to achieve mean square string stability. The  limiting  values for the mean and variance of the tracking error are given by
\begin{align*}
\lim_{i\to \infty}\lim_{k \to \infty} \mu_{\zeta_N}(k) & = 0 ,\\
\lim_{i\to \infty}\lim_{k \to \infty} P_{\zeta_N}(k)  &= \left( \norm{ \frac{S(z)}{M(z)} }_2^2 -1 \right) P_d .
\end{align*} 
\end{theorem}
Theorem \ref{teo:Lpmean_Lqvariance_all} presents conditions for $\mathcal{L}_2$-mean $\mathcal{L}_\infty$-variance string stability of the platoon \eqref{eq:ss_sys_all}, which do not depend on the platoon size $N$ nor the initial conditions of the state. Such conditions are also necessary and sufficient for mean square string stability. We highlight that \eqref{eq:MSS_all} and \eqref{eq:scalability_all} guarantee string stability for the mean and variance separately.
However, this should not be expected for different platooning setups or communication issues, as reported in \cite{govape21,gordon2023mean} for data-lossy channels, where there are cases in which the mean behaves in a string stable fashion but the variance does not. 

\begin{figure*}[t]
	\begin{center}
	    \includegraphics[width=0.95\textwidth, height=7cm]{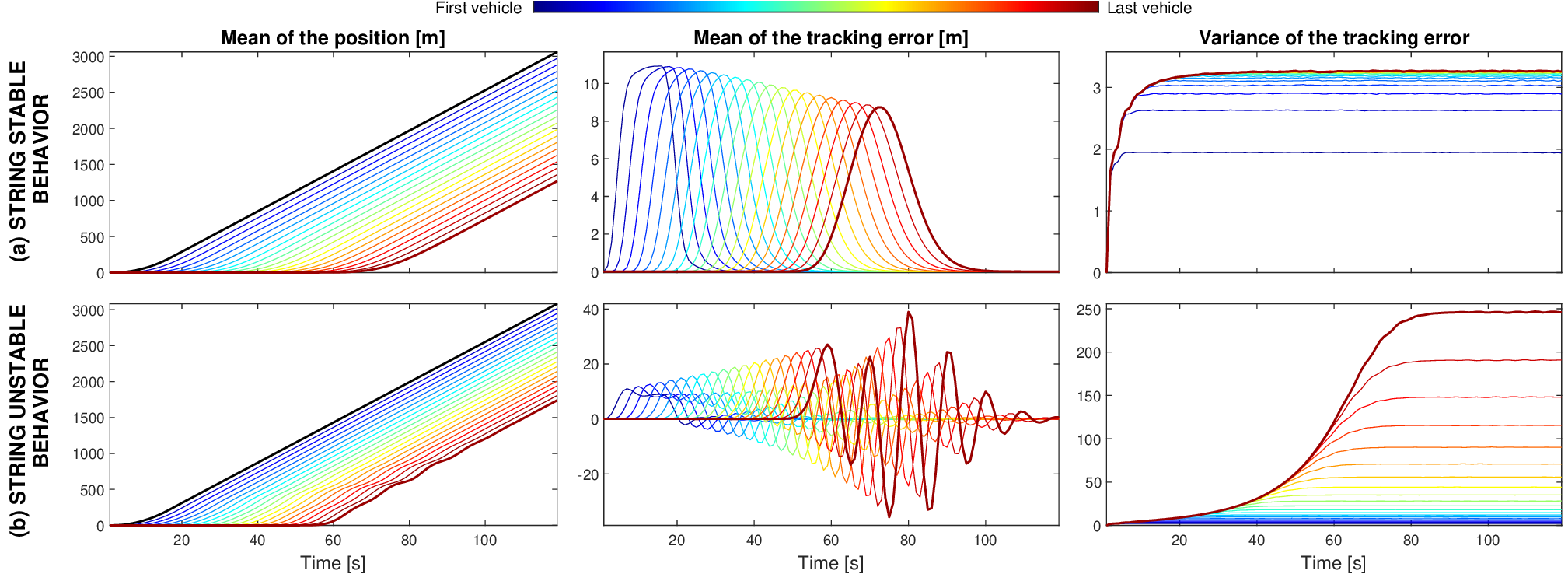}
		\caption{Platoon behavior with zero initial conditions.}
		\label{fig:all_zero_ic}
	\end{center}
\end{figure*}

\begin{figure*}[t]
	\begin{center}
	    \includegraphics[width=0.95\textwidth, height=7cm]{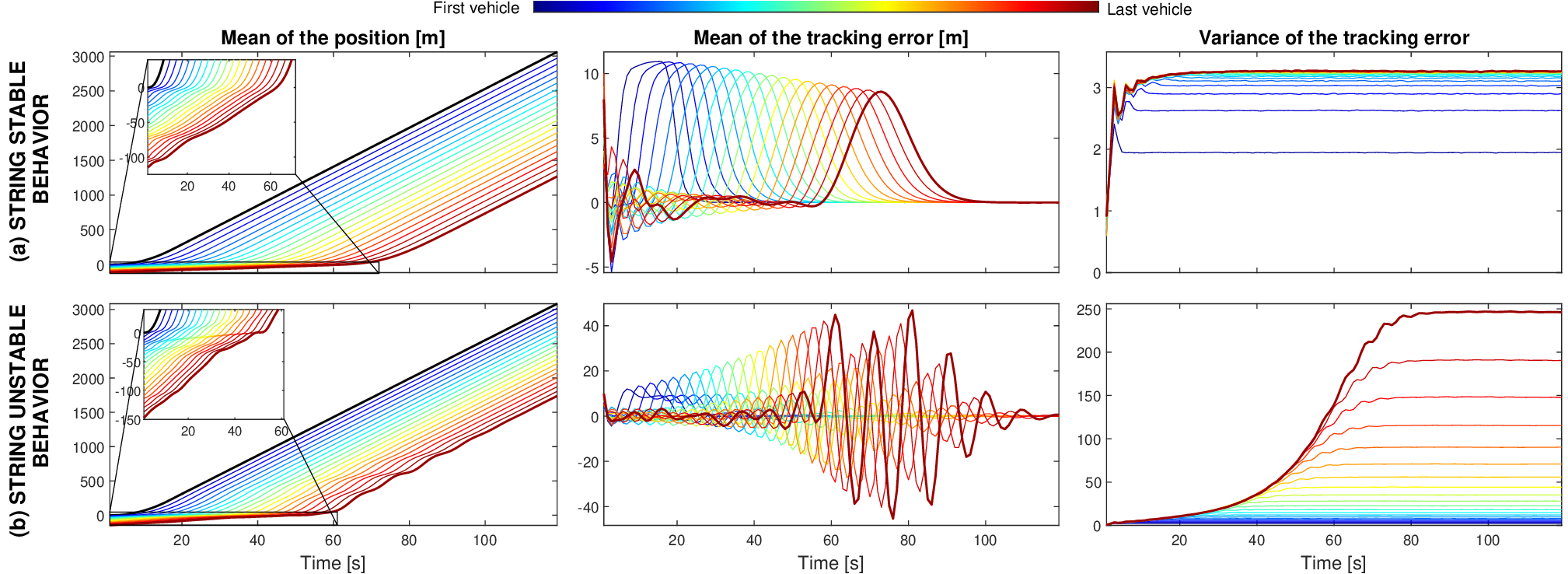}
		\caption{Platoon behavior with initial conditions different from zero.}
		\label{fig:all_dif_ic}
	\end{center}
\end{figure*}


\section{Numerical example}	\label{SEC:example}
In this section, we perform simulations to illustrate our theoretical results. We consider a platoon of 20 homogeneous followers, each with a continuous-time dynamic model given by $G_c(s)=1/s^2$. Hence, the discrete-time model\footnote{zeros at $z=0$ and the gain from the discretization process can be absorbed by the controller.} is given by $G(z)=1/(z-1)^2$. Given $H(z)=(1+h)-hz^{-1}$, we set
\begin{equation*}
     K(z)=\dfrac{(1.35/(1+h))z}{(z+0.89)}, 
\end{equation*}
leading to the closed-loop transfer function
\begin{equation*}
    T(z)=\frac{1.35z}{(1+h)(z+0.89)(z-1)^{2}+1.35(z+hz-h)},
\end{equation*}
which depends on $h$.
To exemplify both stable and unstable string scenarios, we choose two distinct values for the time headway constant. Initially, $h=3.2$ results in string stable behavior, since it can be verified that $\rho(A)=0.5315$ and $|T(e^{j\omega})| < 1, \forall \omega >0$. Conversely, for $h=2.4$, although the convergence in time is satisfied ($\rho(A)=0.6531$), the convergence in the number of vehicles is not achieved since $|T(e^{j\omega})| > 1$, for some $\omega >0$, yielding string instability. We assume the leading vehicle moves with constant speed, the noises $d_i$ are all zero mean with variance $P_{d}=0.6$, and all the agents start from rest. We perform Monte Carlo simulations to obtain the platoon statistics with $1\times 10^6$ realizations. At the top of the subsequent graphics, a colormap is presented to distinguish each agent in the platoon. The first follower is represented by a dark-blue color, while the last follower is depicted in dark-red.

In Fig.~\ref{fig:all_zero_ic} we present a string stable (a) and unstable (b) behavior of the statistics of the tracking error for a platoon with zero initial conditions.  The mean of the tracking error of each vehicle converges to zero in time in both cases since $\rho(A)<1$; however, it is bounded despite the number of vehicles only in the stable case. This shows that mean square stability is a necessary but not sufficient condition for string stability. The behavior of the mean is similar to the one expected for the error in a setting with ideal communication. For the variances, assuming zero initial conditions, we see that the stationary variance of each follower increases with the number of vehicles. In the string stable case, that increment converges to a constant value; while in the string unstable case, it keeps growing unbounded. 
In Fig.~\ref{fig:all_dif_ic}, we repeat the simulation but with non-zero initial conditions.   The string stability properties remain consistent with the previous analysis, with the primary distinction lying in the evolution of transient statistics. 
Though the variance of some vehicles may temporarily exceed that of the last agent (dark-red curve) during the transient behavior, this effect diminishes over time. Steady-state variance values for each vehicle ultimately align with the case of zero initial conditions. 
Our derivations account for the initial conditions, and as the conditions for string stability are independent of them, the overall string stability properties persist.


\begin{figure}[t]
    \begin{center}
    \includegraphics[width=0.99\columnwidth]{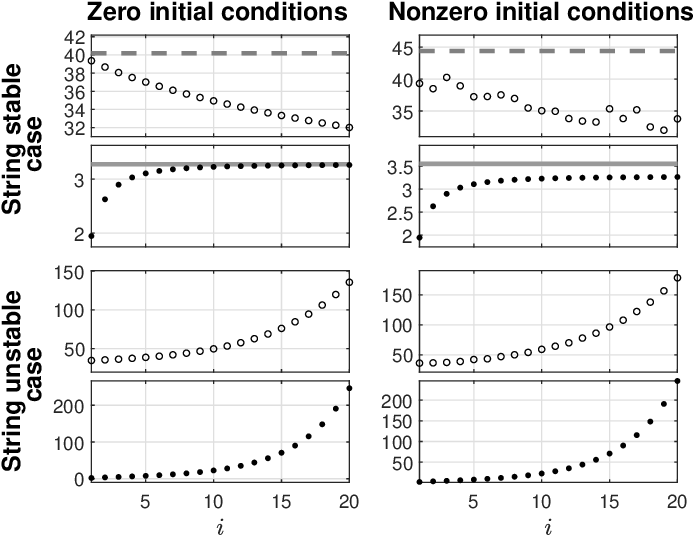}
    \caption{Evolution of the norms and their boundary values, where $\| \mu_{\zeta_i}(k) \|_{\mathcal{L}_2}$ is denoted by $\circ$, the boundary value of the mean obtained with \eqref{eq:L2_norm_mean} is denoted by - -, $\| P_{\zeta_i}(k) \|_{\mathcal{L}_\infty}$ is denoted by $\bullet$ and the boundary value of the variance obtained \eqref{eq:Linf_norm_variance} is denoted by continuous line $-$.} 
    \label{fig:boundary_norms}
    \end{center}                             
\end{figure}


In the context of Definition \ref{def:Lp_Lq_ss}, the mean of errors' curves for our problem enables string stabilization analysis not only for $\mathcal{L}_2$-norm, but also for $\mathcal{L}_1$ and $\mathcal{L}_{\infty}$ norms.
 However, if such means do not converge to zero, $\mathcal{L}_2$-norm analysis is not viable. On the other hand, as the error variance converges to a non-zero value, we focus on the $\mathcal{L}_{\infty}$ norm in such cases instead of the $\mathcal{L}_2$ norm. In different scenarios where the variance converges to zero, such as those presented in \cite{gordon2023mean} involving random data loss, $\mathcal{L}_2$ norm analysis becomes a suitable alternative.\\
In Figure \ref{fig:boundary_norms} we illustrate the evolution of the $\mathcal{L}_2$-norm for the mean, and the $\mathcal{L}_\infty$-norm for the variance of the tracking error.  In the case of string stability, it is observed that the values of the norms converge as the number of vehicles increases, and the bounds obtained according to Definition \ref{def:Lp_Lq_ss} using \eqref{eq:L2_norm_mean} and \eqref{eq:Linf_norm_variance}, are indeed upper bounds. On the contrary, in the case of string instability, such boundary values do not exist, and the norms grow unbounded.

\section{Conclusions and future work}
\label{SEC:conclusion}
The provided stochastic string stability definitions concentrate on the mean and variance of tracking error sequences in interconnected dynamical agent systems. While applicable to various stochastic scenarios, our primary motivation is studying string stability of platoons with additive noise in communication channels.
{In this scenario, we characterize the evolution of the mean and variance of the tracking errors and found that the proposed stochastic string stability requires conditions similar to those in deterministic setups.} 
The findings in this paper lay the groundwork for studying various challenges associated with additive noise channels within the Networked Control Systems framework. 
In future research, we will also explore the impact of diverse channel models on the proposed definitions of string stability. Additionally, we will consider controller synthesis, the study of heterogeneous platoons and the inclusion of saturation in the platoon signals.

\appendix

\section{Proof of Lemma \ref{lem:mean_square_all}} \label{pf:lem_mean_square_all}
 Expressions in \eqref{eq:means_all}  and \eqref{eq:variance_all} are standard results on discrete-time stochastic systems \cite{astrom2012}, where it can also be noted that $\rho(\mathbf{A})<1$ is necessary and sufficient for convergence.
The eigenvalues of the matrix $\mathbf{A}$ are the eigenvalues of the main diagonal blocks, i.e. the eigenvalues of $A$ in \eqref{eq:ith-closedloop} with multiplicity $i$. Therefore, $\rho(\mathbf{A})<1$ if and only if $\rho(A)<1$, which proves \eqref{eq:MSS_all}. The stationary value for the mean is obtained considering that $K(z)G(z)$ has two poles at $z=1$ and thus, each vehicle in the platoon is able to follow ramp signals with zero error in steady state, which implies that $ \displaystyle\lim_{k \to \infty} \zeta_0 (k)=0$ and hence $\mu^{\infty}_{\bm{\zeta}} = 0$. For the stationary variance we use \eqref{eq:variance_all} to note that  $P^{\infty}_{\bm{\zeta}}=\mathbf{C} P^{\infty}_{\bm{\xi}}\mathbf{C}^\top$. Then, applying well-known properties of the $vec$ operator \cite{bernstein2009matrix} it follows that $vec(P^{\infty}_{\bm{\zeta}})=(\mathbf{C} \otimes \mathbf{C}) vec(P^{\infty}_{\bm{\xi}})$, where $vec(P^{\infty}_{\bm{\xi}})=(I- \mathbf{A} \otimes \mathbf{A})^{-1}vec(\Upsilon)$. The proof is completed by applying the inverse operator, $vec^{-1}$. \qed

\section{Proof of Theorem \ref{lem:zdom_scalability_all}} 
\label{pf:lem_zdom_scalability_all}
Applying the expectation operator to \eqref{eq:mu_z_recursive_all}, and considering that
 the platoon is internally stable since $\rho (A)<1$, and the expected value is deterministic, we can apply standard tools from frequency domain analysis and conclude from \eqref{eq:mu_z_recursive_all}
that $\mu_{\zeta_{1}}(e^{j\omega}) = T(e^{j\omega}) \zeta_{0} (e^{j\omega})$ 
 and, for $i>0$, $\mu_{\zeta_{i}}(e^{j\omega}) = T(e^{j\omega}) \mu_{\zeta_{i-1}} (e^{j\omega})$, where we have used the fact that $\mu_{d_{i}}=0$. This proves \eqref{eq:zdom_muz_all}. It is straightforward to note that $|T(e^{j\omega})| \leq 1$ for all $\omega$ is necessary and sufficient for $\mu_{\zeta_{i}}(e^{j\omega})$ to be a bounded sequence in $i$ when $N \rightarrow \infty$, yielding the condition $\norm{T(z)}_\infty \leq 1$. 
 In a similar fashion, we can use \eqref{eq:mu_z_recursive_all} to conclude that
\begin{align*}
   \phi_{\zeta_1}(e^{j\omega}) & =   
        H(e^{j\omega})T(e^{j\omega}) \phi_{d_1}(e^{j\omega})H^*(e^{j\omega})T^*(e^{j\omega})\\
        &= |H(e^{j\omega})T(e^{j\omega})|^2 P_d,
\end{align*}
where we have used the fact that $\zeta_0$ is deterministic and thus $\phi_{\zeta_0}(e^{j\omega})=0$, and also $\phi_{d_i}(e^{j\omega})=P_d$. Similarly, for $i>1$ we have
 \begin{align}
     \phi_{\zeta_{i}}(e^{j\omega})&= |T(e^{j\omega})|^2 \phi_{\zeta_{i-1}}(e^{j\omega}) +|T(e^{j\omega})|^2 \phi_{d_{i-1}}(e^{j\omega}) \notag \\ \label{eq:zeta_rec1}
    &\quad + |H(e^{j\omega}) T(e^{j\omega})|^2 \phi_{d_i}(e^{j\omega})+V(e^{j\omega}),
 \end{align}
where $V(e^{j\omega})$ contains the cross power spectrum density between the three terms in the r.h.s. in \eqref{eq:mu_z_recursive_all}. We note that $d_i$ is uncorrelated with $d_{i-1}$ and $\zeta_{i-1}$ (i.e. $\Phi_{d_{i-1}\zeta_{i-1}}=0$), however $d_{i-1}$ and $\zeta_{i-1}$ are correlated. Hence, from \eqref{eq:mu_z_recursive_all} we have that 
\begin{align}
\nonumber
   V(e^{j\omega}) & = T(e^{j\omega}) \phi_{\zeta_{i-1}d_{i-1}}(e^{j\omega})T^*(e^{j\omega})\\
 \nonumber  
   &=
-P_d|T(e^{j\omega})|^2  \left(H(e^{j\omega})T(e^{j\omega}) \right.\\ \label{eq:zeta_rec2}
&\qquad \qquad  \qquad\left.+H^*(e^{j\omega})T^*(e^{j\omega})\right). 
\end{align}
Equations \eqref{eq:zeta_rec2} and \eqref{eq:zeta_rec1} allow us to write $\phi_{\zeta_{i}}(e^{j\omega})= |T(e^{j\omega})|^2 \phi_{\zeta_{i-1}}(e^{j\omega})+P_dW$,
where
\begin{align}
\nonumber
  W=& |T(e^{j\omega})|^2  \left(1-H(e^{j\omega})T(e^{j\omega}) \right.\\ 
  \label{eq:W1}
&\left.-H^*(e^{j\omega})T^*(e^{j\omega})+|H(e^{j\omega})|^2\right). 
\end{align}
Since $S(e^{j\omega})=1-H(e^{j\omega})T(e^{j\omega})$, we can write
\begin{align*}
\nonumber
  W=& |T(e^{j\omega})|^2  \left(S(e^{j\omega})+S^*(e^{j\omega})-1 \right)+|1-S(e^{j\omega})|^2 \\
=&|T(e^{j\omega})|^2 \ |S(e^{j\omega})|^2+\left( 1-|T(e^{j\omega} \right)|^2)\ |1-S(e^{j\omega})|^2,
\end{align*}
which yields \eqref{eq:zdom_psdz_all}.
Clearly, if $|T(e^{j\omega})|>1$ for $\omega_c\in (0,2\pi]$,  \eqref{eq:zdom_psdz_all} will generate an unbounded sequence in $i$, namely $\phi_{\zeta_i}(e^{j\omega_c})$. When $\omega=0$, the system remains bounded (the power spectral density \eqref{eq:zdom_psdz_all} of the errors generate a constant sequence) since  $W=0$ given that $T(1)=1$ and $S(1)=0$ due to the double integral action of the product $K(z)G(z)$. 
For all $\omega >0$, it is easy to see that a sufficient condition to guarantee boundedness is that $|T(e^{j\omega})| < 1$. On the contrary, when $|T(e^{j\omega_Q})|=1$ for any $\omega_Q>0$, the recursion in  \eqref{eq:zdom_psdz_all} will grow unbounded given that  $|S(e^{j\omega_Q})| = 0$ only for $\omega_Q=0$, since otherwise $S(e^{j\omega_Q})=1-H(e^{j\omega_Q})T(e^{j\omega_Q})=0$ implies $H(e^{j\omega_Q})=1$ which is not possible given the structure of $H(e^{j\omega_Q})$ unless $\omega_Q=0$, which proves necessity.
\qed

\section{Proof of Corollary \ref{cor:stationary_all}} \label{pf:cor:stationary_all}
Employing the induction method, we can systematically substitute the equation \eqref{eq:zdom_psdz_all}, for all $i>1$, and obtain $\phi_{\zeta_i} (e^{j\omega}) = \phi_{\zeta_{i-1}} (e^{j\omega}) +|S(e^{j\omega})T^{i-1}(e^{j\omega})|^2P_d$. Then, from the fact that $P^{\infty}_{\zeta_i} = \frac{1}{2 \pi}\int_{-\pi}^{\pi} \phi_{\zeta_i} (e^{j\omega})  d\omega $, and applying the standard definition of 2-norm for systems, we have 
\begin{equation} \label{eq:pzeta}
  P^{\infty}_{\zeta_i} = 
     P^{\infty}_{\zeta_{i-1}} +\|T^{i-1}(z)S(z)\|_{2}^2P_d.
\end{equation}
This implies $P^{\infty}_{\zeta_i}>P^{\infty}_{\zeta_{i-1}}$ and thus $P^{\infty}_{\zeta_N}$ is the maximum value. Using \eqref{eq:pzeta} recursively, and exploiting 2-norm properties, we note that $P^{\infty}_{\zeta_N}=\|F_{\zeta_N}(z)\|_{2}^2 \; P_d$, where $F_{\zeta_N}$ is as in Corollary \ref{cor:stationary_all}.
To determine the final value when $N \rightarrow \infty$, we notice from \eqref{eq:zdom_psdz_all} that, in such stationary case, the spectrum $\phi_{\zeta_{\infty}} \triangleq \displaystyle\lim_{N\rightarrow \infty} \phi_{\zeta_N}$ satisfies 
	\begin{equation}
	 \phi_{\zeta_{\infty}}(e^{j\omega})=|T(e^{j\omega})|^2\phi_{\zeta_{\infty}}(e^{j\omega})+WP_d,
	\end{equation}
	where $W$ is as in \eqref{eq:W1}. Clearly, $|M(e^{j\omega})|^2 \phi_{\zeta_{\infty}}(e^{j\omega})=WP_d$, with $M(z)$ as in Corollary \ref{cor:stationary_all}. Also, using $S(e^{j\omega})=1-H(e^{j\omega})T(e^{j\omega})$ we can also write $ W= |S(e^{j\omega})|^2+|M(e^{j\omega})|^2 (H(e^{j\omega})T(e^{j\omega})+H^*(e^{j\omega})T^*(e^{j\omega})-1)$.  This allows us to conclude that
	\begin{align}\label{eq:spsta}
\phi_{\zeta_{\infty}}(e^{j\omega})&
	 =\frac{|S(e^{j\omega})|^2}{|M(e^{j\omega})|^2}P_d+\left(H(e^{j\omega})T(e^{j\omega}) \right. \notag \\
	 &\quad + \left. H^*(e^{j\omega})T^*(e^{j\omega})-1\right)P_d.
	\end{align}
Since $\displaystyle\lim_{N\rightarrow \infty} P^{\infty}_{\zeta_N}= \frac{1}{2 \pi}\int_{-\pi}^{\pi} \phi_{\zeta_{\infty}} (e^{j\omega})  d\omega $ we have
\begin{align*}
 \lim_{N\rightarrow \infty} P^{\infty}_{\zeta_N} = & \left(\norm{ {S(z)}/{M(z)}}_2^2 +2 \text{Re} \langle H(z)T(z) , 1 \rangle -1 \right) P_d .
\end{align*}
The proof is completed by noting that, since $T(z)$ is strictly proper, then $H(z)T(z)$ and 1 are orthogonal, and therefore, the inner product $\langle H(z)T(z) , 1 \rangle=0$.\qed 

\section{Proof of Theorem \ref{teo:Lpmean_Lqvariance_all}} 
\label{pf:lem_zdom_scalability_leader}

  \underline{{\it Mean of the tracking error:}} 
    Equation \eqref{eq:means_all} can also be written in terms of the statistics of the initial condition as follows
    \begin{align}
    \label{eq:mu_z_recursive_leader}
    &\mu_{\bm{\zeta}}(k)= \mathbf{C} \mathbf{A}^k \mu_{\bm{\xi}}(0) + \sum_{\ell=1}^{k} \mathbf{C} \mathbf{A}^{k-\ell}\ \mathbf{B}_o \  \zeta_0(\ell-1),
    \end{align}
    where $\mu_{\bm{\xi}}(0)$ is the mean of the initial condition of the state $\bm{\xi}(0)$. 
         Note that $\mu_{\zeta_i}(k)= \eta_i \mu_{\bm{\zeta}}(k)$ with $\eta_i \in \mathbb{R}^{1\times N}$ being a  vector of zeros except for the element at index $i$, which is equal to one. Using the triangle inequality, for each vehicle, we can write
    \begin{align*}
        \norm{\mu_{\zeta_i}(k)}_{\mathcal{L}_2} \leq & \norm{\eta_i \mathbf{C} \mathbf{A}^k \mu_{\bm{\xi}}(0)}_{\mathcal{L}_2}  \\
         &\qquad + \norm{\textstyle\sum_{\ell=1}^{k} \eta_i \mathbf{C} \mathbf{A}^{k-\ell}\ \mathbf{B}_o \ \zeta_0(\ell-1)}_{\mathcal{L}_2}.
    \end{align*}
    Then, for the term associated with the initial condition we compute the norm
    \begin{align}
        \label{eq:beta1_leadera}
        & \norm{\eta_i \mathbf{C} \mathbf{A}^k \mu_{\bm{\xi}}(0)}_{\mathcal{L}_2}^2 \notag \\
        &\hspace{10mm} = \textstyle \sum_{k=0}^{\infty} \eta_i \mathbf{C} \mathbf{A}^k \mu_{\bm{\xi}}(0){\mu_{\bm{\xi}}(0)}^{\top} \mathbf{A}^{k^\top} \mathbf{C}^{\top} \eta_i^{\top} \notag \\
        &\hspace{10mm}=  {\mu_{\bm{\xi}}(0)}^{\top} \textstyle\sum_{k=0}^{\infty}  \left[ \mathbf{A}^{k^\top} \mathbf{C}^{\top} \eta_i^{\top}  \eta_i \mathbf{C} \mathbf{A}^k  \right] \mu_{\bm{\xi}}(0) \notag \\
        &\hspace{4mm}\leq  \norm{\mu_{\bm{\xi}}(0)}_2^2 \  \rho \left( \textstyle \sum_{k=0}^{\infty}  \mathbf{A}^{k^\top} \mathbf{C}^{\top} \eta_i^{\top}  \eta_i \mathbf{C} \mathbf{A}^k \right),
    \end{align}
  where we have used property (a) (see section \ref{SEC:notation}). Hence
    \begin{align}\label{eq:beta1_leader}
        &\norm{\eta_i \mathbf{C} \mathbf{A}^k \mu_{\bm{\xi}}(0)}_{\mathcal{L}_2} \notag \\
        &\hspace{-2mm} \leq \norm{\mu_{\bm{\xi}}(0)}_2 \Big[ \rho \Big(  \textstyle\sum_{k=0}^{\infty} \mathbf{A}^{k^\top} \mathbf{C}^{\top} \eta_i^{\top}  \eta_i \mathbf{C} \mathbf{A}^k \Big) \Big]^{\frac{1}{2}}.
    \end{align} 
    Given the structure of $\eta_i$,  the product $\mathbf{A}^{k^\top} \mathbf{C}^{\top} \eta_i^{\top}  \eta_i \mathbf{C} \mathbf{A}^k$ has rank 1. Also, the specific structure of $\mathbf{A}$ and $\mathbf{C}$ allows to limit the spectral value of $\mathbf{A}^{k^\top} \mathbf{C}^{\top} \eta_i^{\top}  \eta_i \mathbf{C} \mathbf{A}^k$  as $i$ approaches $\infty$.  Given that the spectral radius of the summation of positive semi-definitive matrices, whose spectral radius decreases to zero with $k \rightarrow \infty$ (due to $\rho(A)<1$) is bounded  \cite{bernstein2009matrix}, we conclude that the upper bound in \eqref{eq:beta1_leader} exists and does not depend on $N$.
    
    On the other hand, since $\zeta_0 (k)$ is a deterministic signal and $\mu_{d_i}(k)=0$, the response due to the main input can be treated similarly as in the deterministic case where $\zeta_0(k)$ is a bounded and zero convergent input. From \eqref{eq:zdom_muz_all} and \eqref{eq:mu_z_recursive_all} we notice that $\sum_{\ell=1}^{k} \eta_i \mathbf{C} \mathbf{A}^{k-\ell}\ \mathbf{B}_o \ \zeta_0(\ell-1)$ is the response of $T^{i}$ to the input $\zeta_0$ and thus 
    \begin{align}\label{eq:aux1}
        \norm{\sum_{\ell=1}^{k} \eta_i \mathbf{C} \mathbf{A}^{k-\ell}\ \mathbf{B}_o \ \zeta_0(\ell-1)}_{\mathcal{L}_2} = \norm{T(z)^i}_{\infty} \norm{\zeta_0 (k)}_{\mathcal{L}_2}.
    \end{align}
 Clearly, if $\norm{T(z)}_{\infty} \leq 1$ then \eqref{eq:aux1} is bounded for any type of $\mathcal{L}_2$-norm  bounded reference $\zeta_0(k)$ and $i \in \mathbb{N}$. In this case, since $\mu_{\bm{\nu}}(k) = \zeta_0 (k)$  we have
 \begin{align}
        \label{eq:alpha1_leader}
        \norm{\sum_{\ell=1}^{k} \eta_i \mathbf{C} \mathbf{A}^{k-\ell}\ \mathbf{B}_o \ \zeta_0(\ell-1)}_{\mathcal{L}_2} \leq \norm{T(z)}_{\infty} \norm{\mu_{\bm{\nu}}(k)}_{\mathcal{L}_2}.
    \end{align}
 Since $|T(e^{j0})| = 1$ and $|T(e^{j\omega})| < 1, \forall \omega >0$, implies $\norm{T(z)}_{\infty} \leq 1$, we can conclude that under the aforementioned conditions, the mean of the tracking error satisfies
    \begin{align}
        \label{eq:L2_norm_mean}
        \norm{\mu_{\zeta_i}(k)}_{\mathcal{L}_2} &\leq \alpha_1(\norm{\mu_{\bm{\nu}}(k)}_{\mathcal{L}_2})+\beta_1(\norm{\mu_{\bm{\xi}}(0)}_2),
    \end{align}
    with the class-$\mathcal{K}$ functions $\beta_1$ and $\alpha_1$ determined by \eqref{eq:beta1_leader} and \eqref{eq:alpha1_leader}, respectively.
    
    \underline{{\it Variance of the tracking error:}} We can obtain the recursive expression for the variance of the tracking error as
    \begin{align}
        P_{\zeta_i}(k) = f_i(P_{\bm{\xi}}(0),k) + g_i(P_{\bm{d}},k),
    \end{align}
    where $P_{\bm{\xi}}(0)$ is the variance of the initial condition of the state $\bm{\xi}(0)$,  $g_i(P_{\bm{d}},k)$ is a linear function in $P_{\bm{d}}$ and  $f_i(P_{\bm{\xi}}(0),k)$ is given by    
    \begin{align*}
        f_i(P_{\bm{\xi}}(0),k) &= \eta_i \mathbf{C} \mathbf{A}^{k} P_{\bm{\xi}}(0) {\mathbf{A}^k}^\top \mathbf{C}^\top \eta_i^{\top}, \hspace{0.7cm} \textrm{for}\ k \geq 1. 
    \end{align*}
 The triangle inequality leads to
\begin{align}    
\norm{P_{\zeta_i}(k)}_{\mathcal{L}_\infty}     &\leq \norm{f_i(P_{\bm{\xi}}(0),k)}_{\mathcal{L}_\infty} + \norm{g_i(P_{\bm{d}},k)}_{\mathcal{L}_\infty}.
\end{align}
Exploiting property (a) (see section \ref{SEC:notation}) we write
    \begin{align}
    \norm{ f_i(P_{\bm{\xi}}(0),k)}_{\mathcal{L}_\infty}  &= \max_k \ (\eta_i \mathbf{C} \mathbf{A}^k P_{\bm{\xi}}(0) {\mathbf{A}^k}^\top \mathbf{C}^\top \eta_i^{\top}) \notag \\
    &\hspace{-10mm}\leq \max_k \; \rho ( \mathbf{A}^k P_{\bm{\xi}}(0) {\mathbf{A}^k}^\top ) \  \norm{\eta_i \mathbf{C}}_2.
\end{align}
Since the spectral radius $\rho ( \mathbf{A}^k P_{\bm{\xi}}(0){\mathbf{A}^k}^\top )$ attains its maximum when $k=0$, we can conclude that
\begin{align}
    \label{eq:alpha2_leader}
   \norm{ f_i(P_{\bm{\xi}}(0),k)}_{\mathcal{L}_\infty}\leq & \ \rho (P_{\bm{\xi}}(0))  \norm{\eta_i \mathbf{C}}_2 \notag \\
   =& \ \sigma_{max} (P_{\bm{\xi}}(0))  \norm{C}_2,
\end{align}
where we use property (b) (see section \ref{SEC:notation}) and the fact that $\norm{\eta_i \mathbf{C}}_2^2=\norm{C}_2^2$.

In order to obtain a bound for
$\norm{g_i(P_{\bm{d}},k)}_{\mathcal{L}_\infty}$ we first show that the maximum of $g_i(P_{\bm{d}},k)$  occurs when $k \rightarrow \infty$. For this, we note that $g_i(P_{\bm{d}},k)=P_{{\zeta}_i}(k)$ when $P_{\bm{\xi}}(0)=0$. Given the variance of the noise $P_{\bm{d}}>0$ and assuming $P_{\bm{\xi}}(0)=0$, it is clear from \eqref{eq:P_xi_all} that 
    \begin{align*}
        P_{\bm{\xi}}(1)-P_{\bm{\xi}}(0) = \mathbf{B}_a P_{\bm{d}}{\mathbf{B}_a}^\top \geq 0 .
    \end{align*}
    The next difference is also positive semi-definite since 
    \begin{align*}
     P_{\bm{\xi}}(2)-P_{\bm{\xi}}(1) &= \mathbf{A}  P_{\bm{\xi}}(1) \mathbf{A}^\top + \Upsilon - \mathbf{B}_a P_{\bm{d}} {\mathbf{B}_a}^\top  \\
     &= (\mathbf{A}\mathbf{B}_a+{\mathbf{B}_b})P_{\bm{d}}(\mathbf{A}\mathbf{B}_a+{\mathbf{B}_b})^{\top}  \geq 0. 
    \end{align*} 
    For $k\geq 2$ we have from \eqref{eq:P_xi_all} that
    \begin{align*}
        P_{\bm{\xi}}(k+1)-P_{\bm{\xi}}(k) = \mathbf{A} \left[ P_{\bm{\xi}}(k)-P_{\bm{\xi}}(k-1)  \right] \mathbf{A}^\top. 
    \end{align*}
    Thus, we conclude that $P_{\bm{\xi}}(k+1)-P_{\bm{\xi}}(k) \geq 0, \forall k\geq 0$. Finally, by noting that
    \begin{equation*}
        P_{\bm{\zeta}}(k+1)-P_{\bm{\zeta}}(k) = \mathbf{C} \left[ P_{\bm{\xi}}(k+1)-P_{\bm{\xi}}(k) \right]  \mathbf{C}^{\top},
    \end{equation*}
    it is clear that the difference $P_{\bm{\zeta}}(k+1) - P_{\bm{\zeta}}(k)$ is a positive semi-definite matrix. This also implies that $P_{\zeta_i}(k+1)-P_{\zeta_i}(k)$ is a non-negative number given that each sub matrix of a positive definite matrix is also positive definite \cite{bernstein2009matrix}. Then, when $P_{\bm{\xi}}(0)=0$, $P_{\zeta_i}(k)$ is an increasing function on $k$ and thus, $\sup_k P_{\zeta_i}(k)$ exists 
    
    From Theorem \ref{lem:zdom_scalability_all}  we know that $|T(e^{j\omega})| < 1, \forall \omega >0$, is necessary and sufficient to have a bound for all stationary variances, for any $N \in \mathbb{N}$, and from and Corollary \ref{cor:stationary_all} we know that such bound is given by $P_{\zeta_N}$. Thus, we can conclude that, for any platoon length,
    \begin{align}
    \nonumber
    \norm{g_i(P_{\bm{d}},k)}_{\mathcal{L}_\infty} &\leq
    \norm{g_N(P_{\bm{d}},k)}_{\mathcal{L}_\infty}\\
    \nonumber
     &
     =\norm{P_{\zeta_N}(k)}_{\mathcal{L}_\infty} \quad s.t. \; P_{\bm{\xi}}(0)=0\\
    \label{eq:beta2_leader}
     & =  \left( \norm{ {S(z)}/{M(z)}}_2^2 -1 \right) P_d ,
   \end{align}
where we have used results in Corollary \ref{cor:stationary_all} and the fact that $\norm{P_{\bm{d}}}_{\mathcal{L}_\infty}=P_{d}$. 
Hence, the variance of the tracking error satisfies
    \begin{align}
        \label{eq:Linf_norm_variance}
        \norm{P_{\zeta_i}(k)}_{\mathcal{L}_\infty} &\leq \alpha_2 (\norm{P_{\bm{d}}}_{\mathcal{L}_\infty})+\beta_2(\norm{P_{\bm{\xi}}(0)}_\infty),
    \end{align}
    with the class-$\mathcal{K}$ functions $\alpha_2$ and $\beta_2$ determined by \eqref{eq:alpha2_leader} and \eqref{eq:beta2_leader} respectively.
  The existence of such bounds allows us to conclude that conditions \eqref{eq:MSS_all} and \eqref{eq:scalability_all} are necessary and sufficient for $\mathcal{L}_2$-mean $\mathcal{L}_\infty$-variance string stability. 
  
  From Lemma \ref{lem:mean_square_all} and Theorem \ref{lem:zdom_scalability_all}, it is also clear that conditions \eqref{eq:MSS_all} and \eqref{eq:scalability_all} are also necessary and sufficient for the error statistics to converge to stationary values, which implies mean square string stability. Indeed, it is straightforward to conclude from \eqref{eq:zdom_muz_all}  that $\mu_{\zeta_{N}}(e^{j\omega})$  converges to zero when $N \rightarrow \infty$, while $P^{\infty}_{\zeta_N}$ converges to  the term   in Corollary \ref{cor:stationary_all} \qed


\bibliographystyle{abbrv}
\bibliography{biblio}
\end{document}